\documentclass[%
 reprint,
superscriptaddress,
 amsmath,amssymb,
 aps,
prl,
floatfix,
]{revtex4-2}

\usepackage{graphicx}% Include figure files
\usepackage{dcolumn}% Align table columns on decimal point
\usepackage{bm}% bold math
\usepackage{hyperref}% add hypertext capabilities
\usepackage{url}
\hypersetup{hidelinks}
\usepackage{microtype}
\usepackage[dvipsnames]{xcolor}
\usepackage{makecell}
\usepackage{physics}
\usepackage{upgreek}
\usepackage{amsmath}
\usepackage{siunitx}
\usepackage{bbding}
\usepackage{multirow}
\usepackage{mathrsfs}
\begin{document}

\title{\textbf{Revisiting Phase Stability and Superconductivity in Ca–H Superhydrides with Anharmonic Effects} } 

\author{Wenbo Zhao}
\affiliation{Key Laboratory of Material Simulation Methods and Software of Ministry of Education, College of Physics, Jilin University, Changchun 130012, China}
\affiliation{International Center of Future Science, Jilin University, Changchun 130012, China} 

\author{Zefang Wang}
\affiliation{Key Laboratory of Material Simulation Methods and Software of Ministry of Education, College of Physics, Jilin University, Changchun 130012, China}

\author{Ying Sun}
\email{yings@jlu.edu.cn}
\affiliation{Key Laboratory of Material Simulation Methods and Software of Ministry of Education, College of Physics, Jilin University, Changchun 130012, China}
\affiliation{International Center of Future Science, Jilin University, Changchun 130012, China}

\author{Hefei Li}
\email{lihefei37@jlu.edu.cn}
\affiliation{Key Laboratory of Material Simulation Methods and Software of Ministry of Education, College of Physics, Jilin University, Changchun 130012, China}
\affiliation{State Key Laboratory of Superhard Materials, College of Physics, Jilin University, Changchun 130012, China}

\author{Hanyu Liu}
%\email{hanyuliu@jlu.edu.cn}
\affiliation{Key Laboratory of Material Simulation Methods and Software of Ministry of Education, College of Physics, Jilin University, Changchun 130012, China}
\affiliation{International Center of Future Science, Jilin University, Changchun 130012, China}

\author{Yu Xie}
\email[]{xieyu@jlu.edu.cn}
\affiliation{Key Laboratory of Material Simulation Methods and Software of Ministry of Education, College of Physics, Jilin University, Changchun 130012, China}
\affiliation{Key Laboratory of Physics and Technology for Advanced Batteries of Ministry of Education, College of Physics, Jilin University, Changchun 130012, China}

%\author{Yanming Ma}
%\email[]{mym@jlu.edu.cn}
%\affiliation{Key Laboratory of Material Simulation Methods and Software of Ministry of Education, College of Physics, Jilin University, Changchun 130012, China}
%\affiliation{State Key Laboratory of Superhard Materials, College of Physics, Jilin University, Changchun 130012, China}
%\affiliation{College of Physics, Zhejiang University, Hangzhou 310027, China}
\date{\today}

\begin{abstract}
The prediction of superconductivity above 200 K in CaH$_6$ revolutionized research on hydrogen-rich superconductors, and subsequent experiments have verified this prediction, while unidentified peaks in XRD and the decrease in superconducting temperature upon decompression indicate that unresolved issues remain. In this work, we reconstructed the accurate temperature-pressure phase diagram of the Ca−H system and determined the stability ranges of its candidate superconducting phases by considering anharmonic effects. Our results demonstrate that type-I clathrate Ca$_8$H$_{46-\delta}$ structures become thermodynamically stable at 0 K when anharmonic effects are considered. Notably, we found that the previously predicted CaH$_6$ phase achieves stability above 500 K, underscoring the significant role of temperature and anharmonic effects in stabilizing this intriguing high-pressure phase. Our findings offer insights into the structure and superconducting mechanisms of hydrides.
\end{abstract}
\maketitle
%---------------------------------------------------------------------
\section{1. Introduction}

High-temperature superconductivity has attracted great attention in the field of condensed matter physics\cite{OverviewofSuperconductivityYing-1,OverviewofSuperconductivityYing-2,Overview-Cui,Overview-2021}. Since Ashcroft proposed that metallic hydrides could serve as potential high-temperature superconductors in 2004, hydride-based systems has expanded rapidly\cite{2004}. Although early theoretical calculations predicted superconducting critical temperatures ($T_\mathrm{c}$) of up to $\sim$ 100 K in systems such as SiH, SnH, and PH, these predictions remained unverified experimentally\cite{SiH-1,SiH4,SiH4(H2)2,SnH,PH3-1,PH3-2}. It was not until 2012 that CaH$_6$ was predicted to exhibit superconductivity as high as 215~K under high pressure\cite{CaH6-结构预测}, reigniting interest in hydrogen-rich compounds, and subsequently guiding theoretical and experimental discoveries of record-breaking superconducting materials such as H$_3$S, LaH$_{10}$, and YH$_9$\cite{H3S-1,H3S-2,H3S-3,LaH10-1,LaH10-2,LaH10-3,LaH10-4,LaH10-5,YH6,YH9}, with the recent successful synthesis of the first room-temperature superconductor, LaSc$_2$H$_{24}$\cite{LaScH-t,LaScH-e}, in particular, making hydrogen-rich superconductors a major focus of current superconductivity research.

However, the experimental realization of CaH$_6$ has not been straightforward. Early attempts at synthesis using H$_2$ as the hydrogen source produced only low-hydride phases\cite{cah4-1,cah4-2}. It was not until a decade later, with the introduction of ammonia borane (NH$_3$BH$_3$), that CaH$_6$ was successfully synthesized\cite{CaH6-实验1,CaH6-实验2}. X-ray diffraction (XRD) confirmed its structure to be consistent with theoretical predictions, and $T_\mathrm{c}$ as high as 215~K was observed. Yet, additional unidentified diffraction peaks were also detected in the XRD patterns. Moreover, although theory predicts that the $T_\mathrm{c}$ should increase during decompression, the experiment instead shows that it stays nearly constant from 200 to 165 GPa and then decreases sharply below 165 GPa. These results suggest that the phase landscape of the Ca–H system may be more complex than previously anticipated.

Recently, more refined structural predictions have revealed a new cubic phase, Ca$_8$H$_{46}$\cite{Ca8H46-Miao,Ca8H46-An}, which adopts the type-I clathrate structure initially synthesized in Ba$_8$Si$_{46}$\cite{Ba8Si46}. With the inclusion of Ca$_8$H$_{46}$ in the Ca–H phase diagram, this phase becomes thermodynamically stable, whereas CaH$_6$ becomes metastable. The calculated $T_\mathrm{c}$ is also close to the experimental results. Moreover, hydrogen vacancy structures (such as Ca$_8$H$_{45}$ and Ca$_8$H$_{44}$) lead to structural distortions and a significant reduction in $T_\mathrm{c}$, consistent with the experimentally observed trend of $T_\mathrm{c}$ suppression upon decompression. These observations imply that Ca$_8$H$_{46-\delta}$ may also be present among the synthesized phases. Therefore, determining the temperature–pressure stability ranges of these candidate superconducting phases is a great topic.

We employed machine learning potential based large-scale calculations to construct a temperature–pressure phase diagram of the Ca–H system that includes anharmonic effects, and identified the stability regions of potentially superconducting phases. The results show that Ca$_8$H$_{46-\delta}$ is thermodynamically stable at 0~K, while CaH$_6$ becomes stable at high temperatures ($\sim$ 500~K). A reduction in hydrogen content significantly suppresses superconductivity in both CaH$_{6-\delta}$ and Ca$_8$H$_{46-\delta}$, which accounts for the observed decrease of $T_\mathrm{c}$ upon decompression. This work not only clarifies the structural origin of superconductivity under different conditions, but also highlights the critical role of lattice anharmonicity in hydrogen-rich superconductors. The established framework provides a consistent explanation for the theoretical and experimental in the Ca–H system and offers theoretical guidance for the future design of stable high-$T_\mathrm{c}$ hydride materials.

\section{2. Results}
\subsection{Harmonic Temperature–Pressure Phase Diagram}
To compute the temperature–pressure phase diagram of the Ca–H system, we first performed an comprehensive machine-learning accelerated structure search for Ca$_8$H$_x$ ($32 \le x \le 48$) at 130 GPa and constructed the 0~K harmonic phase diagram in the 130–200 GPa range. Consistent with previous findings\cite{Ca8H46-Miao,Ca8H46-An}, Ca$_8$H$_{44}$, Ca$_8$H$_{45}$, and Ca$_8$H$_{46}$ become sequentially the most thermodynamically stable phases as pressure increases, culminating in the stabilization of Ca$_8H_{46}$ above 180 GPa. In contrast, CaH$_{6}$ remains metastable throughout, with an energy above the convex hull ($E_\mathrm{hull}$) of 20–26 meV/atom, which decreases with increasing pressure (Table~S1).
\begin{figure}[htbp]
    \centering
    \includegraphics[width=1\linewidth]{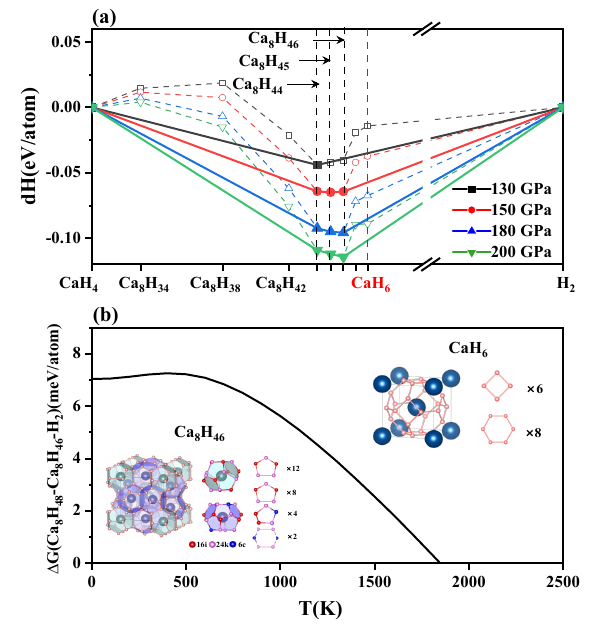}  
    \caption{(a) Formation enthalpies of Ca–H compounds (130–200 GPa) relative to CaH$_4$ and H$_2$ at the harmonic level. Solid and open symbols denote stable and metastable phases, respectively. (b) Temperature-dependent Gibbs free energy difference between Ca$_8$H$_{48}$ and Ca$_8$H$_{46}$ + H$_2$ at the harmonic level.}
\end{figure}

Given hydrogen's high vibrational frequencies, zero-point energy (ZPE) corrections are critical for accurately assessing hydride stability. To this end, phonon calculations were carried out for the candidate Ca–H structures. These calculations revealed that Ca$_8$H$_{46}$ is dynamically stable only above 200 GPa, which is stricter than previously reported\cite{Ca8H46-Miao,Ca8H46-An}, possibly due to the use of more stringent computational parameters (Supplementary Section 3). The stability threshold is reduced to 130 GPa when anharmonic effects are considered.

Therefore, ZPE corrections were applied only at 200 GPa. At this pressure, the energy of CaH$_6$ drops to just 7 meV/atom above the convex hull at 0 K. To further evaluate its finite-temperature stability, we calculated the Gibbs free energies, which show that CaH$_6$ becomes thermodynamically stable at temperatures above roughly 1900 K. However, this estimate is based on solid H$_2$ as the reference state, while H$_2$ is liquid above 800 K at 200 GPa. Consequently, the actual stability temperature of CaH$_6$ is likely even higher, and may exceed the experimental synthesis temperature ($\sim$2000 K). These findings highlight the limitations of harmonic approximations in explaining CaH$_6$ formation under experimental conditions. To gain a more comprehensive understanding of phase stability and transition features in the Ca–H system, we further include anharmonic effects in our study.

\subsection{Anharmonic Temperature–Pressure Phase Diagram}

We performed anharmonic free energy calculations on key Ca$_8$H$_{44-48}$ structures from the harmonic phase diagram using our previously developed SSCHA–ACNN method\cite{sscha-acnn}, which combines the stochastic self-consistent harmonic approximation (SSCHA)\cite{sscha-2013,sscha-2017,sscha-2018,SSCHA-2021} with an attention-based neural network potential (ACNN)\cite{ACNN}. This approach significantly reduces computational cost, enabling anharmonic calculations for complex, multi-atom structures (see Supplementary Section 4 for details). 

As shown in Fig. 2, including anharmonic effects has a noticeable impact on the predicted phase stability. At 0~K, the Ca$_8$H$_{44}$–Ca$_8$H$_{46}$ phases are now positioned directly on the convex hull throughout the 130–200 GPa range, with Ca$_8$H$_{46}$ consistently identified as the most energetically favorable composition. While CaH$_{6}$ remains slightly metastable, it lies only 7–9 meV/atom above the hull, comparable to that obtained with harmonic ZPE (Table~S1).

The anharmonic phonon spectra confirm the dynamical stability of all Ca$_8$H$_{44}$–Ca$_8$H$_{46}$ phases down to 130 GPa (Fig. 2b–e and Fig. S5-S8). This stability stems from a systematic renormalization of the hydrogen-related manifold where high-frequency modes soften and low-frequency modes harden, accompanied by a modest lattice expansion without symmetry breaking.

More importantly, the temperature-dependent anharmonic Gibbs free energies reveal that CaH$_{6}$ becomes thermodynamically stable between 100 and 400 K. This re
\begin{figure*}[t]
    \centering
    \includegraphics[width=1\textwidth]{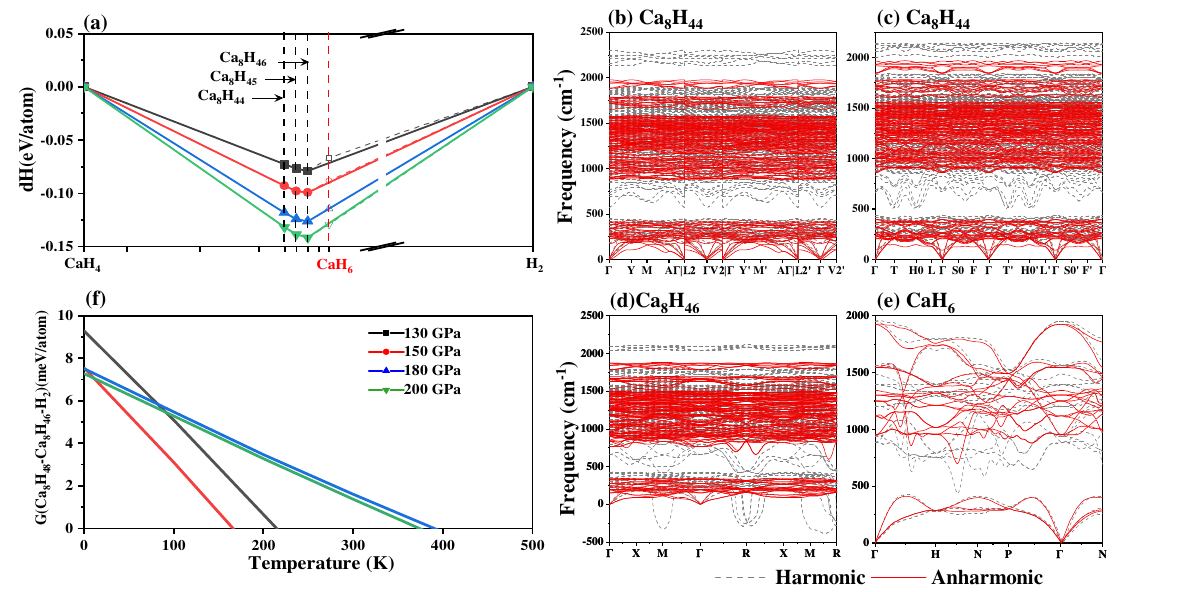}  
    \caption{(a) Anharmonic formation enthalpies of Ca–H compounds (130–200 GPa) relative to CaH$_4$ and H$_2$. Solid and open symbols denote stable and metastable phases, respectively. (b–e) Anharmonic phonon spectra for (b) Ca$_8$H$_{44}$, (c) Ca$_8$H$_{45}$, (d) Ca$_8$H$_{46}$, and (e) CaH$_6$ at 150 GPa. Red solid and gray dashed lines represent anharmonic and harmonic results, respectively. (f) Temperature-dependent Gibbs free energy difference between between Ca$_8$H$_{48}$ and Ca$_8$H$_{46}$ + H$_2$ at the anharmonic level.}
\end{figure*}
sult notably refines the phase stability evolution, demonstrating that CaH$_{6}$ is accessible under far more moderate thermal conditions than the harmonic approximation suggests. This enhanced accessibility facilitates the capture of CaH$_{6}$ during high-temperature synthesis, whereas at lower temperatures, Ca$_8$H$_{46}$ emerges as the energetically preferred ground state.

\subsection{Superconductivity}

To further understand superconducting behavior, we examined the structural and electronic properties of Ca$_8$H$_{42–48}$ compositions. Structure predictions reveal that most compositions within this range can stabilize in two distinct hydrogen cages, CaH$_6$-type and Ca$_8$H$_{46}$-type, denoted as CaH$_{6-\delta}$ and Ca$_8$H$_{46-\delta}$, respectively. The only exception is Ca$_8$H$_{48}$ (CaH$_6$), where excess hydrogen suppresses formation of the Ca$_8$H$_{46}$-type clathrate structure, and can only stabilize the CaH$_6$-type cage. All compositions favor the Ca$_8$H$_{46}$-type cage energetically, except for Ca$_8$H$_{48}$ (CaH$_6$) (Table~S4). Notably, as the hydrogen content increases, the energy of the CaH$_6$-type framework gradually decreases, narrowing the energetic gap between the two distinct hydrogen cages.
\begin{figure}[htbp]
    \centering
    \includegraphics[width=1\linewidth]{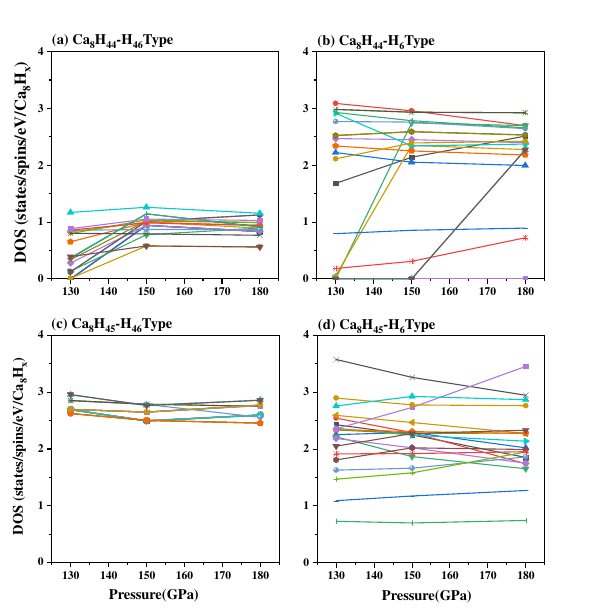}  
    \caption{Pressure dependence of the density of states at the Fermi level, $N(E_\mathrm{F})$, for Ca$_8$H$_{44}$ and Ca$_8$H$_{45}$ in Ca$_8$H$_{46-\delta}$ and CaH$_{6-\delta}$ structural types.}
\end{figure}

\begin{figure}[htbp]
    \centering
    \includegraphics[width=1\linewidth]{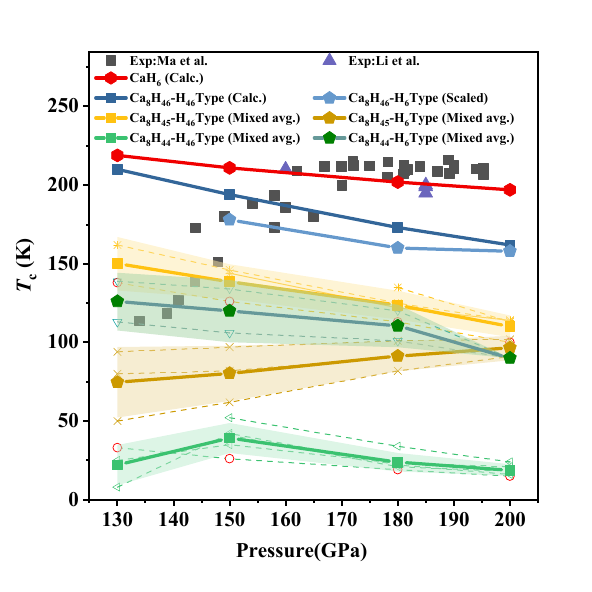}  
%    \caption{Pressure dependence of anharmonic $T_\mathrm{c}$ for CaH$_6$ and Ca$_8$H$_{44-46}$ in Ca$_8$H$_{46-\delta}$ and CaH$_{6-\delta}$ structural types. Open symbols/dashed lines and solid symbols/solid lines represent individual and averaged results, respectively. Shaded regions denote estimated error ranges. Experimental data are from Ma \textit{et al.}~\cite{CaH6-实验2} (black squares) and Li \textit{et al.}~\cite{CaH6-实验1} (purple triangle).}
    \caption{Pressure dependence of anharmonic $T_\mathrm{c}$ for CaH$_6$ and Ca$_8$H$_{44-46}$ in Ca$_8$H$_{46-\delta}$ and CaH$_{6-\delta}$ structural types. For CaH$_6$ and Ca$_8$H$_{46}$, data sources are indicated in parentheses: (Calc.) denotes explicit anharmonic superconductivity calculations and (Scaled) denotes values obtained from scaling corrections. For Ca$_8$H$_{45}$ and Ca$_8$H$_{44}$, open symbols/dashed lines show the individual results: the first point (highlighted by a red circle) is from explicit anharmonic calculations, whereas the remaining points are obtained from scaling estimates (Scaled). Solid symbols/solid lines represent the corresponding averaged results. Shaded regions denote estimated error ranges. Experimental data are from Ma \textit{et al.} (black squares) and Li \textit{et al.} (purple triangle).}
\end{figure}

We selected representative low-enthalpy structures derived from these two cage types and systematically analyzed their electronic density of states (see Supplementary Section 6). The results show that while cage geometry and structural symmetry affect the electronic structure, stoichiometry plays the dominant role. As hydrogen content decreases, the density of states at the Fermi level ($N(E_\mathrm{F})$) gradually drops for both CaH$_{6-\delta}$ and Ca$_8$H$_{46-\delta}$ (Fig.~S8). Some structures of Ca$_8$H$_{44}$ may even become insulating and require compression to regain metallicity under low pressure\cite{band-pressure,LaH5,RbS2}. The relevant pressure window and a brief discussion of its origin are provided in Supplementary Section~5. Notably, Ca$_8$H$_{48}$ (CaH$_6$) shows a lower $N(E_\mathrm{F})$ than Ca$_8$H$_{46}$ despite having more hydrogen, while the fully occupied Pm$\bar{3}$n–Ca$_8$H$_{46}$ structure exhibits the highest $N(E_\mathrm{F})$.

%Due to computational limitations, anharmonic superconductivity calculations were performed only for the stable phases at each composition (Table~S8). 
Due to computational limitations, anharmonic superconductivity calculations were performed only for the stable phases at each composition; $T_\mathrm{c}$ is evaluated with $\mu^*=0.10$ unless otherwise stated, and values for $\mu^*=0.13$ and $0.15$ are provided in Table~S8 of the Supplementary Information. The results show a consistent trend across compositions: structural symmetry remains unchanged, lattice volume slightly expands, the frequency range of hydrogen-related optical phonons becomes significantly compressed, and some phonon modes that strongly contribute to electron–phonon coupling (EPC) are weakened, leading to a consequent decrease in the $T_\mathrm{c}$ (Fig.~S15-S16).

Guided by this trend, we performed harmonic calculations on representative Ca$_8$H$_{46-\delta}$ and CaH$_{6-\delta}$ cages and applied scaling corrections to estimate $T_\mathrm{c}$ (see Supplementary Section 7 for details). The results show that CaH$_6$ exhibits the highest predicted $T_\mathrm{c}$, closely matching experimental observations under high-temperature and high-pressure synthesis (Fig.~5). The $T_\mathrm{c}$ of Pm$\bar{3}$n–Ca$_8$H$_{46}$ is lower than that of CaH$_6$, with the difference increasing to approximately 30~K as pressure increases, which differs from previous reports at harmonic level\cite{Ca8H46-An}. The convergence tests of superconducting parameters are shown in Supplementary Section 3. Ca$_8$H$_{46}$ with a CaH$_6$-type clathrate structure exhibits an even lower $T_\mathrm{c}$. The deviation from the trend observed in $N(E_\mathrm{F})$ may be attributed to the high phonon frequencies and strong electron–phonon coupling in CaH$_6$, further supporting it as the origin of the high-$T_\mathrm{c}$ superconducting phase reported in earlier experiments.

Meanwhile, a consistent trend: CaH$_6$ $\textgreater$ Ca$_8$H$_{46}$ $\textgreater$ Ca$_8$H$_{45}$ $\textgreater$ Ca$_8$H$_{44}$, is observed in both Ca$_8$H$_{46-\delta}$ and CaH$_{6-\delta}$ cages, at both harmonic and anharmonic levels. This progression explains the rapid suppression of superconductivity observed experimentally upon decompression.

%---------------------------------------------------------------------
\section{3. Conclusion}

Our work clarifies the stability ranges of all potentially superconducting phases in the Ca–H system across pressure and temperature. We identify Ca$_8$H$_{46-\delta}$ as the thermodynamically stable phase at low temperatures, while CaH$_{6-\delta}$ becomes accessible at high temperatures ($\sim  500$ K). The observed decrease in $T_\mathrm{c}$ upon decompression is primarily attributed to a reduction in hydrogen content. Our findings clearly explain the experimentally observed emergence of high-$T_\mathrm{c}$ and low-$T_\mathrm{c}$ phases and their temperature and pressure dependence. Moreover, highlight the critical role of temperature and anharmonic effects in stabilizing high-pressure superconducting phases.

\section{Data availability statement}
The SSCHA code (https://github.com/SSCHAcode/python-sscha) is open source and is based on the GNU General Public License v3.0. The ACNN code is available from Y. X. upon reasonable request. All data in the paper are available from the corresponding author upon request.

\section{Acknowledgment}
This work was supported by the National Natural Science Foundation of China (Grant No. 12374008, 12022408, 12304013, 12374009, 12074138, 22131006, 52288102, and 52090024), the Interdisciplinary Integration and Innovation Project of JLU, Fundamental Research Funds for the Central Universities and the Program for JLU Science and Technology Innovative Research Team (JLUSTIRT), open project from state key laboratory of superhard materials (No. 202408). 

\section{Author contribution statement}
Y.X. designed and supervised the project. W.Z. performed most of the calculations and drafted the manuscript. Y.S. contributed to figure design and assisted with writing and revising the manuscript. Z.W. performed part of the harmonic calculations. H.Li, and H.Liu contributed to manuscript revision and discussion. All authors have read and approved the final version of the manuscript.
\nocite{*}

\bibliography{references}% Produces the bibliography via BibTeX.

\end{document}